\documentclass[preprint,aps,prd,nofootinbib,tightenlines]{revtex4}

\usepackage{amsmath}
\usepackage{graphicx}

\def\mres{m_{\rm res}}

\def\CSB{\raise0.4ex\hbox{$\chi$}SB}
\def\chpt{\raise0.4ex\hbox{$\chi$}PT}
\def\schpt{S\raise0.4ex\hbox{$\chi$}PT}

\def\tr{{\rm tr}}

\def\LQCD{\Lambda_{\rm QCD}}

\def\Dslash{{\rm D}\!\!\!\!/\,}

\def\spose#1{\hbox to 0pt{#1\hss}}
\def\ltapprox{\mathrel{\spose{\lower 3pt\hbox{$\mathchar"218$}}
 \raise 2.0pt\hbox{$\mathchar"13C$}}}
\def\gtapprox{\mathrel{\spose{\lower 3pt\hbox{$\mathchar"218$}}
 \raise 2.0pt\hbox{$\mathchar"13E$}}}
\def\inapprox{\mathrel{\spose{\lower 3pt\hbox{$\mathchar"218$}}
 \raise 2.0pt\hbox{$\mathchar"232$}}}

\begin{document}

\title{Future of Chiral Extrapolations
with Domain Wall Fermions\footnote{%
Extended version of talk given
at workshop on ``Domain Wall Fermions at Ten Years'',
Riken-BNL Research Center, March 15-17, 2007.}
}

\author{Stephen R. Sharpe}
\email[]{sharpe@phys.washington.edu}
\affiliation{%
Physics Department, University of Washington,
Seattle, WA 98195-1560, USA}

\date{\today}

\begin{abstract}
I discuss the constraints on the lattice spacing, $a$,
the quark masses, $m$, the box size, $L$,
and particularly the residual mass, $\mres$, such that one can successfully
calculate phenomenologically interesting quantities
using Domain Wall fermions (DWF).
The constraints on $a$, $m$, and $L$ are largely common with
other improved fermion discretizations, and I emphasize that
the improved chiral symmetry of DWF does not remove the need
for simulations with a significant range of lattice parameters.
Concerning $\mres$, I compare the analysis of chiral symmetry
breaking to that with Wilson fermions, emphasizing
that DWF are better than simply Wilson fermions with 
each chiral symmetry breaking effect reduced by a common factor.
I then discuss the impact of non-zero $\mres$ both on generic hadronic
quantities, and on matrix elements which involve mixing with
lower dimension operators. 
\end{abstract}


\maketitle


\section{Introduction}

Simulations with domain-wall fermions (DWF)~\cite{Kaplan,DWF}
have made impressive
progress over the last few years. Algorithmic improvements
(particularly the work of Ref.~\cite{RHMC})
have allowed simulations with $2+1$ flavors of dynamical fermions to begin
in earnest.
Precision results for 
important quantities have been obtained, e.g. $B_K$~\cite{DWFBK}
and the $K\to\pi$ form factor~\cite{DWFKtopi}, albeit at
a single lattice spacing. 
Thus it is a good time to look ahead a few years and
study the strengths and limitations of the approach~\cite{DWFat10}.

The key attraction of DWF compared to Wilson-like fermions
is the reduction in the size of chiral symmetry breaking.
This comes at the cost of adding an extra flavor-like dimension,
with concomitant increase in computational cost. 
The length of the extra dimension, $L_5$, is chosen
as a compromise between speeding up the simulations (by reducing $L_5$)
and decreasing the size of chiral symmetry breaking (by increasing $L_5$).
The main focus of this talk is to study the impact of the
residual chiral symmetry breaking (\CSB) on calculations of phenomenologically
important quantities.

I cannot resist setting things up as a choice between three
alternatives:
\begin{itemize}
	\item 
{\bf D}ARN GOOD: 
No practical barrier to calculating
{\bf some} quantities of interest (spectrum, $B_K$ \dots)
with desired precision in next 5 years.

\item {\bf  W}ONDERFUL:
 No practical barrier to calculating
{\bf many} quantities of interest (spectrum, $B_K$, 
some $K\to\pi\pi$ matrix elements,
 \dots) with desired precision in next 5 years.

\item {\bf F}ANTASTIC: 
No practical barrier to calculating
{\bf most} quantities of interest (spectrum, $B_K$, $\epsilon'/\epsilon$,
condensate \dots)
with desired precision in next 5 years.
\end{itemize}
There are no bad (or ugly~\cite{SSlat06}) options here---i.e. it
is expected that the theory has the correct continuum limit.
The only theoretical concern is whether the effective four-dimensional
action is local, and this can be checked numerically,
as has been done for present simulations~\cite{RBC07}.
There is, however, the practical issue of how far calculations can be
pushed given available computational resources.
What is needed is an ensemble of
lattice ensembles with a range of
lattice parameters ($a$, $m_\ell$, $m_s$, and box size $L$) 
comparable to that of the present MILC staggered-fermion ensemble.
This would allow validation of the methods by
comparing to experimentally measured quantities, independent predictions of those
quantities which have been accurately 
calculated with staggered fermions,\footnote{%
Here I am injecting my opinion that it is plausible that
rooted-staggered fermions give rise to the correct continuum
limit, based on the considerations and work of others summarized
in Refs.~\cite{SSlat06,BGSlat06}.}
and first controlled calculations of the quantities which are difficult to obtain
using staggered or Wilson-like fermions. It is the latter quantities
that are my focus here.

The outline of this talk is as follows. In the
next section I briefly recall the range of simulation parameters,
in particular the light quark masses, that will likely
be needed to extract precision results. This is to emphasize
the importance of creating the ensemble of ensembles mentioned
above, and to bring out a couple of points that are sometimes forgotten.
I then turn to the core of this talk, namely an analysis of
the impact of the residual \CSB\ on various
important quantities: pion properties, the quark condensate, electroweak
matrix elements with and without power divergences, and
the pion electromagnetic mass splitting.
I also discuss the extent to which one can think of DWF as
{\em ``Wilson-fermions-lite,''} i.e. Wilson fermions with the
\CSB\ reduced (very significantly) in magnitude.
I close by attempting a choice from the three options given above.

I should stress that 
I am mainly pulling together work in the literature
and attempting to provide a uniform discussion,
although I am adding some new observations along the way.

\section{How small do $a$, $m$ and $1/L$ need to be?}

In this section I make a few remarks on
the range of simulation parameters
that one should aim for.\footnote{%
I do not discuss constraints on $L_5$ in this section---these 
will be discussed at length below. For the discussion in this section,
and in particular for the ``basic'' quantities under consideration,
I assume that $L_5$ is fixed at a large
enough value that residual \CSB\ effects are smaller than
other errors.}
This range obviously depends on
the quantity under consideration and the desired precision. What
I have in mind here are ``basic'' quantities (spectrum, decay constants,
matrix elements without subtractions, \dots) and percent-level accuracy.
Nothing I say here is special to DWF, but conversely one needs to
keep in mind that the positive features of DWF do not exempt them
from these standard requirements.

\subsection{How small does $a$ need to be?}

Although residual \CSB\ introduces discretization
errors linear in the lattice spacing $a$ (at least
for generic quantities, as discussed further below),
the dominant such error for most quantities is of relative
size ${\cal O}(a\Lambda)^2$. In other words, DWF are (almost) automatically
$O(a)$ improved. This puts DWF simulations in 
essentially the same situation
as almost all the other large-scale simulations,
i.e. those using overlap fermions,
non-perturbatively $O(a)$ improved Wilson fermions,
twisted-mass fermions, or staggered fermions.

The scale characterizing discretization errors should be comparable to
$\LQCD\approx 0.3\,$GeV,
and I start by taking $\Lambda=0.1-0.5\;$GeV.
With this, the relative size of the leading discretization errors is
\begin{equation}
(a\Lambda)^2= (0.003-0.06) \left(\frac{a}{0.1\ {\rm fm}}\right)^2 \,.
\label{eq:discerrors}
\end{equation}
Since one has, or aims to have, two or more lattice spacings,
this leading error should be largely removed by extrapolation to $a=0$,
leaving a significantly smaller residue.
Thus it is not unreasonable to think that percent-level accuracy
should be attainable with $a_{\rm min} \approx 0.1\;$fm.
This is consistent with the RBC/UKQCD plans described at this meeting,
which involve moving from the present runs with $a\approx 0.12\,$fm to
runs (already underway) at $a\approx 0.09\,$fm.

Is this assessment correct? 
Are simulations with $a_{\rm min} \approx 0.1\;$fm sufficient?
I raise this issue because there is evidence that smaller lattice
spacings are needed for improved Wilson fermions, despite the
fact that the leading discretization errors are of the same 
parametric size as for DWF.\footnote{%
Improved staggered fermions are also pushing to smaller
lattice spacings---$a\approx 0.06\;$fm at present, and possibly
$a\approx 0.045\;$fm in the future.
The need for smaller lattice spacings is greater for staggered
fermions, however, because of the need to disentangle taste
breaking from the chiral limit.}
Some of the evidence is collected in
Sommer's recent lectures~\cite{Sommernara}.
He quotes the following examples, all using non-perturbatively
improved Wilson fermions:
\begin{itemize}
\item 
10-15\% discretization errors at $a=0.1\,$fm seen in quenched $m_q$ and $f_K$
\cite{Garden99}.
These errors are larger than the range above, perhaps indicating
that one should use scales as large as $\Lambda=750\;$MeV.
\item
Similar effects seen for $N_f=2$~\cite{Sommer03}.
\item
Large differences between results using different definitions
of $Z_A$ (which should agree in the continuum limit)
for $a\gtapprox 0.1\,$fm.\cite{DellaMorte05}.
\end{itemize}
Perhaps because of this,
recent high-precision calculations with Wilson-like
fermions have worked with lattice spacings down to
$a_{\rm min} \approx 0.05\,$fm~\cite{Luscher,QCDSF}.

One should also note, however, that improved
Wilson fermions do differ from DWF. In particular,
the leading sub-dominant error for improved Wilson fermions
is proportional to $a^3$, an error which is much suppressed for DWF
because it is proportional to the residual $\chi$SB.
It could be, therefore, that discretization errors are smaller
with DWF, as has been observed for some quantities
using overlap fermions~\cite{WittigWennekers05,Liu06}.
Nevertheless, it seems prudent to expect larger discretization
errors than the estimate in (\ref{eq:discerrors}) suggests,
at least for some quantities, and thus that one may need to work
at significantly smaller lattice spacings than $0.1\,$fm.

\subsection{How small does {$m_q$} need to be?}

\begin{figure}[htb]
\includegraphics[width=7cm]{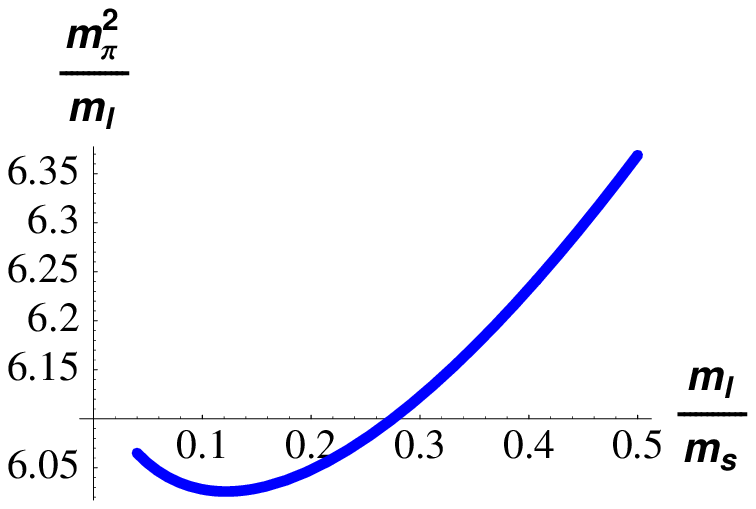}
\hfill
\includegraphics[width=7cm]{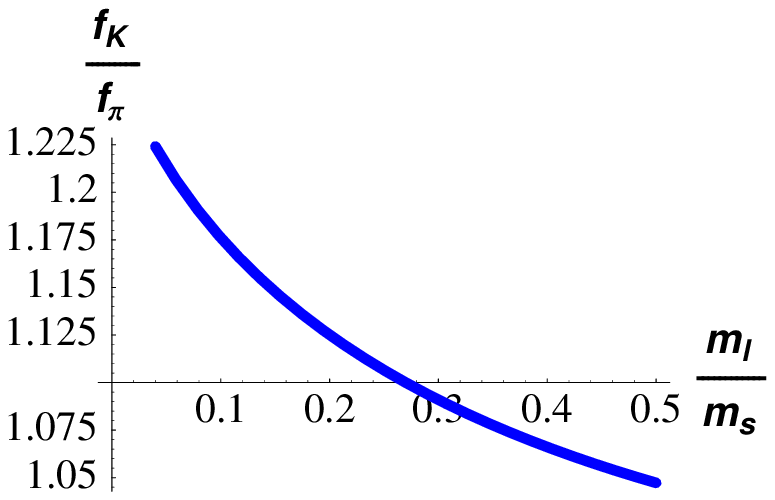}
\caption{Examples of curvature due to the presence of chiral logarithms.
$m_s$ is held fixed, while $m_\ell$ is varied down to its physical value.
See text for more discussion.}
\label{fig:chilog}
\end{figure}

Concerning the light quark masses, it probably bears repeating
that extrapolations to the physical value of the average light quark mass,
$m_\ell \equiv (m_u+m_d)/2 \approx m_s^{\rm phys}/27$,
require that one work at small enough $m_\ell$ to be able
to see and fit the non-analytic contributions.
Examples of such ``chiral logarithms''
(taken from Ref.~\cite{SSnara}, where further discussion may be found)  
are shown in Fig.~\ref{fig:chilog}.
Here I have plotted the results of continuum \chpt\ taking
$m_s^{\rm phys}=0.08\;$GeV, $f\approx0.09\;$GeV, and, for the 
Gasser-Leutwyler coefficients, $L_5=1.45\times 10^{-3}$,
$L_8=10^{-3}$ and $L_4=L_6=0$ at a scale $\mu=m_\rho$~\cite{Bijnens04}.
These curves are indicative of what one should find when
doing unquenched chiral extrapolations (and not precise predictions).

My conclusion is that one needs to have accurate results
down to $m_\ell/m_s^{\rm phys}\approx 0.1$ for precision extrapolations.
This is of course a generic conclusion---individual quantities or
ratios may have smaller non-analytic terms and be more easily
extrapolated~\cite{Becirevicratios}.
It is also ``old news''---the MILC collaboration has worked down
to such masses and found that they do allow precision 
extrapolations~\cite{MILCfpifK},
and other simulations are aiming at, or have already attained, such
small masses.
But perhaps it is good to keep being reminded of it!

\subsection{How large does $L$ need to be?}

\begin{figure}[htb]
\includegraphics[width=10cm]{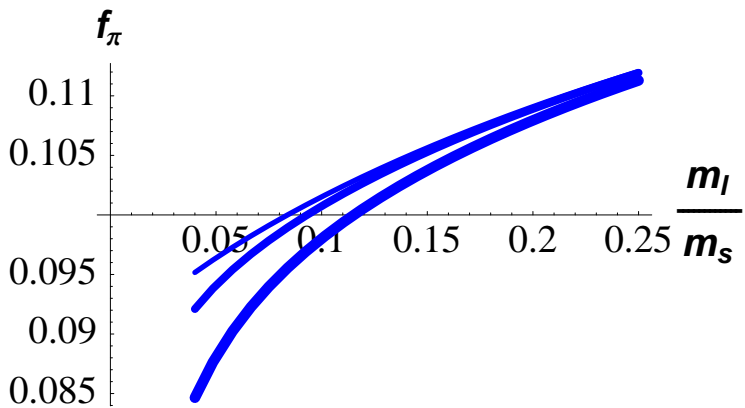}
\caption{Example of finite volume effects, as predicted
by one-loop $\chi$PT, assuming $a=0.1\,$fm, and taking
$L=2.4\,$fm (thick line--at bottom), $3.2\,$fm (medium line-middle)
and $\infty$ (thin line-top). Parameters as for other figures,
except that $f=0.08\;$GeV.
See text for more discussion.}
\label{fig:finiteV}
\end{figure}

Given the cost of simulating with DWF, work to date has
been on relatively small lattices with a physical size of
about $2\,$fm. Here I want to emphasize the importance
of using larger lattices, given that finite volume effects can be substantial.
I illustrate this in Fig.~\ref{fig:finiteV}, which shows that
at $m_\ell/m_s=0.1$, finite volume shifts are at the few percent level.
This is well-known, and indeed generation
of $L=3\,$fm DWF lattices at $a\approx 0.12\,$fm
is underway in order to study, and reduce, finite volume effects.

What is perhaps less well-known is that one-loop \chpt\ does
not, in general, give an accurate estimate of volume corrections---one
needs to include (at least the dominant part of) the two-loop 
contributions~\cite{Colangelo05}. 
To say it differently, one-loop results give, in general, a good indication
of the magnitude of the effect but are not accurate enough to
do a correction accurate at the percent level. 
Since two-loop calculations are hard, 
and are likely to be available for only a few quantities, a
conservative conclusion is that one must work in volumes large
enough so that the one-loop correction is smaller than the desired
accuracy. This pushes one to even larger volumes, although how large
depends on the quantity and the desired precision.

\section{Implications of residual chiral symmetry breaking}

Domain-wall fermions 
work by having left- and right-handed quarks (with small four-dimensional
momentum) bound to different defects, in this case the two walls at either
end of the fifth dimension.
Chiral symmetry---the separate rotation of left- and right-handed quark fields---is
broken by the overlap of the bound-state ``wavefunctions''.
This rough description can be made more precise using a transfer-matrix formalism
to describe propagation in the fifth dimension~\cite{NN,FurmanShamir}.
The result is that, for the interactions of the low-momentum four-dimensional
quark fields (i.e. those with $p \ll 1/a$), the size of
\CSB\ can be parameterized by the 
residual mass (the various definitions of will be discussed below)
whose dependence on $L_5$ is schematically~\cite{GSSloc,RBC07}:
\begin{equation}
\mres \sim \frac{e^{-\alpha L_5}}{L_5} + \frac{\rho(0)}{L_5} \,.
\label{eq:mresform}
\end{equation}
I take $\mres$ to be {\em dimensionless} in this talk, primarily because
it simplifies the comparison with Wilson-like fermions.
Some intuition concerning the form of $\mres$ will be useful in the following,
so I give a brief discussion.

The first term on the r.h.s. corresponds to the picture of modes
exponentially localized on the walls, which thus have an overlap falling
exponentially with $L_5$.
In the transfer matrix language, this term arises from
delocalized eigenmodes, which occur for eigenvalues above
the ``mobility-edge'' $\alpha$.
In perturbation theory $\alpha\sim O(1)$, and
depends on the negative mass used
in the fifth dimension and on the
choice of action~\cite{AokiTaniguchi,ShamirPT}. 

The power-law term in (\ref{eq:mresform}) is not visible in perturbation
theory, but arises in the transfer-matrix analysis from near-zero modes
of the corresponding Hamiltonian, of which $\rho(0)$ is the density per unit
four-volume. A key observation is that the corresponding (four-dimensional)
eigenfunctions are localized at the scale of the lattice 
spacing~\cite{localization}.
Crudely speaking, a left-handed quark
propagating on one wall has, generically, only an exponentially damped
amplitude to ``tunnel'' to the other wall (and thus become right-handed),
but occasionally encounters a near-zero mode, which
can transport it to the other wall with an
essentially unsuppressed amplitude.
There are, however, constraints on simultaneous ``transports'' of 
multiple fermions which will be important later~\cite{Christlat05}.
As for $\alpha$,
the zero-mode density $\rho(0)$ can be altered, and in particular reduced,
by tuning the gauge and DWF actions. 

It is important to keep in mind that both delocalized and localized
eigenmodes of the transfer matrix
lead to a local effect in the long-distance
four-dimensional effective theory, and in particular to a local
mass term at leading order in the Symanzik expansion.
Here local means exponentially
localized at the lattice scale. For the localized eigenmodes, this
is automatic. For the delocalized eigenmodes, however, this must arise
from the collective effect of the modes in the vicinity of the
mobility edge. Heuristically, the argument for locality is that, if the mobility
edge is of $O(1/a)$, i.e. of order its perturbative value, then,
in the viewpoint in which the fifth dimension is a flavor
index, one has $L_5$ four-dimensional flavors
coupled with a mass matrix with generic entries of $O(1/a)$. 
Diagonalizing this matrix leads to the single light Dirac fermion
and $L_5-1$ heavy fermions with masses of $O(1/a)$. Integrating
the latter out will lead to a local four dimensional theory.\footnote{%
The locality of the effective four-dimensional Dirac operator for
DWF has been shown to hold
(independent of $L_5$), under the assumption of sufficiently small
lattice field strengths (which forbids zero modes of the Wilson kernel), 
or in the presence of an isolated zero-mode~\cite{Kikukawa}.
These are the same conditions under which locality of the overlap
Dirac operator has been established~\cite{HJL}.
As for overlap fermions, the result
is not directly applicable to actual DWF simulations,
in which no constraint on lattice field strengths is applied, and which have a non-zero
density of near-zero modes. Thus in practice one must rely on the mobility-edge
discussion of Ref.~\cite{localization}, and check numerically for locality,
as in Ref.~\cite{RBC07}.}
For discussion of the locality of the residual
mass term in the transfer-matrix formalism see Refs.~\cite{Christlat05,RBC07}. 
It can also be seen in perturbation theory~\cite{AokiTaniguchi,ShamirPT}.

I will use the label $\mres$ in various ways in this talk.
It has a fundamental, theoretical definition, as a coefficient in 
the Symanzik effective Lagrangian [eq.~(\ref{eq:Symanzik}) below], 
but with this definition it is not directly calculable. 
It is defined in practice using 
eq.~(\ref{eq:mresdef}) below, but this includes discretization errors
(absent in the fundamental $\mres$)
which make it (weakly) quark-mass dependent rather than a constant. 
I follow recent usage
and call the calculable quantity $\mres'(m)$~\cite{Antonio06,Allton07}.
Finally, I will also use $\mres$ generically to mean
``a chiral symmetry breaking quantity with similar magnitude to
the fundamental value''. In this generic usage the dependence
on $L_5$ has the same form as in eq.~(\ref{eq:mresform}), but
the coefficients of the two terms differ
from those for the fundamental $\mres$.\footnote{%
Note, however, that the coefficient of the exponential, $\alpha$,
does not change (for a given action), since 
it is determined by the mobility edge.
I am grateful to Yigal Shamir for emphasizing this to me.}
The main reason for this generic usage 
is that the Symanzik Lagrangian is invalid inside UV divergent loop
integrals---such loops give rise to quantities with a similar
magnitude and form to the fundamental $\mres$, but which are
numerically  different.
Furthermore, I will be making order-of-magnitude estimates, involving
unknown coefficients, so a precise numerical value for $\mres$ will not
be needed.

What matters, then, is the approximate magnitude
of $\mres$ in present and upcoming simulations.
A lot of work has clarified that if the coupling is too strong
(i.e. if $a$ is too large) then $\alpha$ becomes too small
and chiral symmetry is lost.
Happily, one can avoid this problem by moving towards weak coupling,
and further mitigate it by tuning the action.
It is also noteworthy that $\rho(0)$, the coefficient of the
power-law term in $\mres$, is expected to drop very rapidly 
as $a\to 0$~\cite{rhozerovanishing}.
For present $2+1$ flavor dynamical simulations
at $a\approx 0.12\;$fm ($1/a\approx 1.6\,$GeV), with $L_5=16$,
and the Iwasaki gauge action,
one has ${\mres=0.003}$~\cite{coarseDWF}.
In physical units this is $\mres/a \approx 5\, {\rm MeV}$, i.e.
slightly larger than $m_\ell^{\rm phys}$.
Moving to $a\approx 0.09\,$fm ($1/a\approx 2.2\,$GeV)
at the same $L_5$, one finds
a further reduction, with $\mres\ltapprox0.001$
($\mres/a \ltapprox 2\, {\rm MeV}$)~\cite{Mawhinneyhere}, 
although the precise value
is not yet known.\footnote{%
The very recent study of Ref.~\cite{RBC07} allows one to
determine the relative contribution of the
exponential and power-law contributions to $\mres'$
[see eq.~(\protect\ref{eq:mresform})].
One finds that, for $L_5=16$, these terms
are close to equal  at $1/a\approx 1.6\,$GeV,
while at $1/a\approx 2.2\,$GeV the exponential term
is larger by a factor of about 6.}
Eventually, as $a$ decreases
at fixed $L_5$, one expects from perturbation theory
that the dimensionless $\mres$
asymptotes to a constant up to logarithms, 
so that $\mres/a$ will start to increase. 
Fortunately, simulations have not yet entered this asymptotic regime.

The bottom line is that in the next couple of years one will have
$\mres \approx a m_\ell^{\rm phys}$, and the question is whether
this is small enough to calculate various quantities of interest.

\subsection{Comparing Wilson and DW fermion analyses of chiral symmetry breaking}

I begin by recalling the classic analysis of \CSB\
for (unimproved) Wilson fermions~\cite{Bochetal},
since it will serve as the template for the subsequent discussion of DWF.
Explicit breaking of chiral symmetry leads to a violation of
the PCAC relation
\begin{equation}
\partial_\mu A_\mu^b = 2 m P^b + a X^b \,,
\label{eq:Boch1}
\end{equation}
where $A_\mu^b$ is a particular discretization of
the flavor non-singlet axial current,
$P^b$ the ultra-local pseudoscalar density, and $m$ the quark
mass (taking degenerate quarks for simplicity).
$a X^b$ arises from the Wilson-term in the fermion action, and
thus at tree level has the explicit factor of $a$ as shown.
This is canceled in matrix elements by factors of $1/a$
from UV divergences, corresponding to the dimension-5 operator $X^b$ mixing
with lower-dimensional operators. To understand the impact of the
extra term, then, one must determine the operators of dimension four or less 
with which $X^b$ can mix. The answer is that the lattice symmetries allow
\begin{equation}
a X^b \sim - 2 \frac{m_c}{a} P^b - (Z_A-1) \partial_\mu A_\mu^b + O(a)\,,
\label{eq:Boch2}
\end{equation}
where $\sim$ indicates equivalence in on-shell matrix elements
(so that $X^b$ does not come into contact with other operators)
with $p \ll 1/a$ (so that $O(ap)$ corrections are small).

The first term on the r.h.s. of (\ref{eq:Boch2}) can be absorbed
by an additive renormalization of the quark mass,
with $m_q^{\rm phys} \propto m-m_c/a$,
while the second leads to a renormalization of the axial current.
After these renormalizations one regains the PCAC relation 
up to corrections which are truly of $O(a)$, i.e.
$\partial_\mu (Z_A A_\mu^b) - 2 (m-m_c/a) P^b \sim O(a)$.
To implement this, however, one needs two conditions
to determine the two constants $m_c$ and $Z_A$.
One approach is to use
the vacuum to pion matrix element of eq.~(\ref{eq:Boch2}),
\begin{equation}
\langle 0| a X^b|\pi\rangle = 
- 2 (m_c/a) \langle 0| P^b |\pi\rangle
- (Z_A-1) \langle 0|\partial_\mu A_\mu^b |\pi\rangle\,,
\label{eq:Boch3}
\end{equation}
which determines one linear combination
of the two constants,\footnote{%
The condition is usually written in the equivalent way
$\langle 0| Z_A \partial_\mu A_\mu^b - 2 (m\!-\!m_c/a) P^b|\pi\rangle=0$,
which determines $Z_A^{-1}(m-m_c/a)$. This form makes clear
that it is not necessary to determine $X^b$ explicitly,
and that one can implement the procedure for any desired forms of
$A_\mu^b$ and $P^b$ [e.g. with the ultra-local axial current rather
than the simply local form appearing in eq.~(\protect\ref{eq:Boch1})]. 
I write the condition as given in the text, eq.~(\protect\ref{eq:Boch3}),
as it makes the connection to the subsequent DWF analysis more
transparent.}
while the second condition is obtained by enforcing
the normalization of Ward identities relating 2-point and
3-point functions~\cite{Bochetal}.
The details 
are not important here---what matters is that there are
two constants to determine 
so one needs two conditions.

Now consider the corresponding analysis for DWF.
The axial current is that generated by dividing the fifth dimension
into two halves and doing opposite flavor rotations in each,
and has a lattice divergence~\cite{FurmanShamir}
\begin{equation}
\partial_\mu A_\mu^b = 2 m J_5^b + 2J_{5q}^b \,.
\label{eq:DWFPCAC}
\end{equation}
Here $J_5$ is the pseudoscalar density built out of fields on the
two walls,
while $J_{5q}$ is similar in form to $J_5$ but ``lives'' half-way
across the fifth dimension.
The $J_{5q}$ term is due to the explicit \CSB\ caused by 
using finite $L_5$.
Thus the situation is formally similar to that with Wilson fermions,
with $J_5$ corresponding to $P$ (up to differences in normalization), 
and $J_{5q}$ to $aX$.\footnote{%
That $J_{5q}$ does not contain an overall factor of $a$, unlike $aX$,
is technically correct, but perhaps misleading. 
For finite $L_5$, $J_{5q}$ does not vanish in the classical
continuum limit, whereas $aX$ does. On the other hand, as noted above,
UV divergences imply that matrix elements of $aX$ do not have an
overall factor of $a$, but rather vanish logarithmically (as powers of
$g(a)^2$) in the continuum limit. In practice this amounts to
very little suppression of $aX$, and is a much smaller effect than
the exponential suppression of $J_{5q}$.}

In the standard analysis~\cite{RBC00}
one now argues that, for low-energy matrix elements, the symmetry
breaking term can mix only with the pseudoscalar density:
\begin{equation}
J_{5q}^b \sim \frac{\mres}{a} J_5^b + O(a) \,.
\label{eq:J5qtoJ5}
\end{equation}
Furthermore, because $J_{5q}$ resides in the middle of the fifth dimension,
this mixing is suppressed by the decay of the physical modes attached
to the walls, so that, unlike the Wilson case, $\mres \ll 1$.
The result of this discussion
is that one has only an additive renormalization of
the quark mass.
The single renormalization constant can be determined by a single
condition, with the standard choice 
being~\cite{BlumSoni,Blumlat98,Aokietal99,AliKhan00,RBC00}:
\begin{equation}
 \frac{\mres'}{a} \equiv \frac{\langle0| J_{5q}^b |\pi\rangle}
                        {\langle0| J_5^b |\pi\rangle} \,.
\label{eq:mresdef}
\end{equation}
In other words, one enforces (\ref{eq:J5qtoJ5})
as an exact relation in the vacuum to pion matrix element, 
including $O(a)$ terms. The inclusion of these terms causes
$\mres'$ to depend on $m$ (as we will see in more detail below),
so it now picks up a prime to indicate that it is not a 
constant~\cite{Antonio06,Allton07}.
It is natural to define a shifted quark mass
\begin{equation}
m_q(m) = m + \mres'(m)/a = 
\frac{\langle0| \partial_\mu A_\mu^b |\pi\rangle}
     {2\langle0| J_5^b |\pi\rangle}
\,,
\label{eq:mqdef}
\end{equation}
which in Wilson-fermion parlance is the ``Ward identity mass''.
Since the r.h.s. is proportional to $m_\pi^2$, it follows that
the pion mass necessarily vanishes
when $m_q=0$. This occurs when $m = - \mres'(m)/a$.

Comparing the Wilson and DWF analyses, one immediately wonders
why the latter is more simple, involving one parameter rather than two.
Should eq.~(\ref{eq:J5qtoJ5}) not be replaced by
\begin{equation}
J_{5q}^b \sim \frac{\mres}{a} J_5^b 
- \frac{(Z_A - 1)}{2} \partial_\mu A_\mu^b + O(a) \,?
\label{eq:J5qtoJ5true}
\end{equation}
The answer is 
yes---the enumeration of operators
is just as for Wilson fermions and eq.~(\ref{eq:J5qtoJ5true}) is
the correct form.
In practice, however, the standard analysis, 
and eq.~(\ref{eq:J5qtoJ5}) in particular,
are not invalidated. This is
because the $\partial_\mu A_\mu$ term
is highly suppressed, $Z_A-1 \propto \mres^2\ltapprox 10^{-5}$,
and can be ignored in practice.

Why is $Z_A-1$ suppressed by two powers of $\mres$?
The answer can be deduced from work in the literature. 
Since $Z_V=1$ for any $L_5$, $Z_A-1=Z_A-Z_V$ can be calculated
in perturbation theory (or using non-perturbative renormalization)
by determining the size of the matrix elements of the left-handed current
(which is associated with one wall) 
between right-handed quarks
(attached to the other wall).
Thus one needs {\em two} crossings of the fifth-dimension,
one in each direction, to obtain
a contribution to $Z_A-1$. (This is the analog of the result
with Wilson fermions that $Z_A-1$ is proportional to $r^2$,
with $r$ the coefficient of the Wilson term.)
In perturbation theory each crossing is
exponentially suppressed and proportional to $\mres$~\cite{AokiTaniguchi}.
Using the transfer matrix argument one reaches the same conclusion
in a non-perturbative context~\cite{Christlat05}---in particular,
the structure of the near-zero modes does not allow one such mode
to cause two simultaneous crossings of the type required
for $Z_A-1$.\footnote{%
The argument of Ref.~\cite{Christlat05} is made in the context
of mixing of left-left four-fermion operators, but it applies
also to $Z_A-Z_V$. The observation that
the interaction induced by near zero-modes is ``'t Hooft-like'',
and thus highly constrained,
was made in Ref.~\cite{sailing}.}

In summary, the fact that some of the terms 
which enter into a Wilson fermion analysis (here $Z_A-1$) 
receive an extra suppression
is an example of how DWF are better than 
``Wilson fermions with an overall suppression of \CSB\ effects''.

\subsection{Implications of $\mres\ne0$ in pseudo-Goldstone boson sector}

The analysis so far has absorbed the leading effect of \CSB\
into a shift in the quark mass. In this section I recall
how this can be seen directly in the Symanzik effective
Lagrangian, and extend the discussion to sub-leading effects
suppressed by powers of $a$.
I consider the 
application to the pseudo-Goldstone boson (PGB) sector, 
and investigate the relative size of
errors resulting from \CSB\ (which turn out to be
linear in $a$) compared to (chirally unsuppressed) errors 
of ${\cal O}(a^2)$. For simplicity of presentation I assume degenerate
quarks---the generalization to non-degenerate quarks is straightforward.

The Symanzik effective Lagrangian~\cite{Symanzik} 
for DWF has exactly the same form as for Wilson fermions~\cite{WilsonLeff},
since the symmetries are the same. For on-shell quantities one 
has~\cite{RBCquenchedME}
\begin{equation}
{\cal L}_{\rm Sym.} =
\bar q (\Dslash+Z_m m) q + \frac{Z_m m_{\rm res}}{a} \bar qq
+ a c_5 \bar q (\sigma\cdot F) q + O(a^2) \,,
\label{eq:Symanzik}
\end{equation}
where $q$ are quark fields in a regularized continuum theory
in which factors of the lattice spacing are now explicit.\footnote{%
This description hides a technical
difference between the Wilson and DWF analyses
related to the dependence on $a$. This point was discussed
at the workshop, and here I offer my opinion.
The Symanzik action for Wilson fermions is of the standard form,
with the dependence on $a$ given by the explicit
factors of $a^n$ together with an implicit
logarithmic dependence contained in the coefficients of the terms.
In the DWF case, there is an additional, more complicated,
$a$ dependence contained in $\mres$, eq.~(\protect\ref{eq:mresform}).
In particular, as noted above, $\rho(0)$ has a rapid, possibly exponential,
dependence on $a$, and is a non-perturbative quantity.
Does this invalidate the application of the Symanzik expansion to DWF?
I think not, for two reasons---one general and one specific to DWF.
First, although the Symanzik expansion was originally derived in
perturbation theory, where one finds powers of $a$ up to logarithms,
it can be thought of as an effective field theory (EFT), a concept which is
not tied to perturbation theory. There is no need for the parameters
in the EFT to have a simple dependence on $a$. 
Second, the presence of the additional parameter $L_5$ allows for 
the appearance of a more
complicated $a$ dependence. Indeed, when studying DWF in perturbation theory
one is effectively constructing terms in the EFT, and one finds a
complicated dependence on $a$, with, for example, the exponent
$\alpha$ in eq.~(\protect\ref{eq:mresform}) 
being a power series in $\alpha_S(a)$~\cite{ShamirPT}.}
Thus $q$ are proportional (at leading order in $a$) to the usual
boundary quark fields of DWF.
The factor of $Z_m$ is needed since $m$ is the standard lattice
quark mass [i.e. the same quantity as in eq.~(\ref{eq:DWFPCAC})]
and thus must be matched to continuum quantities.
\CSB\ allows the extra, UV divergent mass term,
whose coefficient provides the fundamental definition of $\mres$.
\CSB\ also allows the Pauli-term, the coefficient of which is
has the same form
as $\mres$, eq.~(\ref{eq:mresform}), and is thus also highly suppressed.
Note, however, that this is only a correspondence of magnitudes,
and that the detailed dependence of $\mres$ and $c_5$ on $L_5$ will differ,
because the ratio of the coefficients
of the exponential and power-law terms differ.

We are now in a position to determine the form of the relationship between $\mres$
as defined in (\ref{eq:Symanzik}) and $\mres'$ from (\ref{eq:mresdef}).
To do this we first match ${\cal L}_{\rm Sym.}$ onto the
chiral effective theory, following the method of Ref.~\cite{SharpeSingleton}.
This leads to
\begin{eqnarray}
{\cal L}_{\chi} &=&
\frac{f^2}{4} \tr(\partial_\mu \Sigma \partial_\mu \Sigma^\dagger)
- \frac{f^2 B}{2} \tr (M_q \Sigma + \Sigma^\dagger M_q) + \dots 
\label{eq:chiralL}\\
M_q &=& m + \frac{m_{\rm res}}{a} + a c_5 \Lambda^2 \,,
\label{eq:chiralMq}
\end{eqnarray}
where $\Sigma$ is the usual non-linear PGB field,
$f$ the leading order pion decay constant (normalized so that
$f_\pi=93\,$MeV,
$B$ is proportional to the condensate (and absorbs $Z_m$),
and the dots in (\ref{eq:chiralL}) indicate terms of higher
order in $m$, $m_{\rm res}$, $a$ and derivatives.
The important result contained in eqs.~(\ref{eq:chiralL},\ref{eq:chiralMq})
is that the PGB masses come not only from the quark masses
in ${\cal L}_{\rm Sym.}$, but also receive a contribution
from the Pauli term.
This is because the Pauli term has the same chiral transformation
properties as the quark mass terms. The size of the Pauli-term
contribution is not, however, known precisely---both
because $c_5$ is unknown and because the mapping to $\chpt$
introduces the scale $\Lambda$ which, while of
${\cal O}(\LQCD)$, is not known otherwise.\footnote{%
While it is redundant in practice to have the
product of two unknown quantities, $c_5$ and $\Lambda^2$, I keep
both since they originate from different 
steps in the matching of the lattice theory onto $\chpt$.}

The net result is that the $O(a)$ term from ${\cal L}_{\rm Sym.}$ is
completely absorbed into $M_q$, at leading order in \chpt. 
Since {$m_\pi^2 \propto M_q$}, and recalling from above
that $m_\pi$ vanishes when $m_q=m+\mres'/a=0$, one finds
\begin{equation}
M_q=m_q \quad \Rightarrow\quad \mres'/a  = \mres/a + c_5 a\Lambda^2\,,
\label{eq:mres'vsmres}
\end{equation}
up to higher order chiral corrections, and terms of size $\mres^2$.
In other words, when we use the standard definition of
$\mres'$ we are automatically including the contribution to
the pion masses (and scattering amplitudes, etc.) from the Pauli-term.
We do not need to introduce an additional \CSB\ term of size
$c_5 a\Lambda^2 \sim {\mres a \Lambda^2}$ into the
leading order {${{\cal L}_\chi}$}.\footnote{%
As noted in Ref.~\cite{SharpeSingleton}, the same automatic
$O(a)$ improvement of the leading order chiral Lagrangian holds
for Wilson fermions. The difference here is that the term
which is excluded is already suppressed.}

This automatic improvement does not work for higher order terms
in the chiral expansion.
The operators {$\bar q \sigma\cdot F q$}
and {$\bar qq$} are different, and their matrix elements
not in general proportional.
The impact of this difference has been worked out to next-to-leading
order for Wilson fermions~\cite{WChPTNLO}, and can be carried
over to DWF with appropriate extra suppressions of \CSB\ terms.
A mnemonic which allows one to see the size of the 
resulting discretization errors is to take the continuum result
and make the substitution
$m_q \to m_q + \mres a \Lambda^2$ (except in the leading
order chiral Lagrangian as discussed above). 
Two examples are
\begin{equation}
\frac{m_\pi^2}{m_q} \sim f_\pi \sim
{\rm const.} \left[
1 + {\cal O}(m_q/\Lambda) + {\cal O}(\mres a \Lambda)
+ {\cal O}(a^2\Lambda^2) +\dots \right]\,.
\label{eq:genericerrors}
\end{equation}
Thus \CSB\ gives rise to multiplicative discretization errors
of relative size $\mres a \Lambda$---i.e. 
suppressed by $\mres$ compared to the
size expected for unimproved Wilson fermions.
Similar corrections are present in all hadronic quantities.

Are these errors important in practice, given present values of $\mres$?
I think the answer is clearly no. They are numerically tiny and
much smaller than the chirally unsuppressed $O(a^2)$ terms.
To make this concrete, consider the situation with present 
DWF simulations at $1/a=1.6\,$GeV, and use $\Lambda=0.1-0.5\,$GeV.
Then
\begin{equation}
\mres a \Lambda \approx 0.003 \left(\frac{0.1-0.5\ {\rm GeV}}{1.6\ {\rm GeV}}\right)
\approx (2-9)\times 10^{-4} \ll (a \Lambda)^2 \approx 0.004-0.1 \,.
\label{eq:genericestimate}
\end{equation}
For percent-level accuracy, the \CSB-induced errors can be ignored.

\subsection{Mass dependence of $\mres'$}

Numerical results for $\mres'$ 
show a weak linear dependence on the bare quark mass $m$ 
(see, e.g., Refs.~\cite{Antonio06,Allton07}).
This is understood in the literature to be a discretization error, but here I want
to determine more precisely its origin and form.
Since $\mres'$ is defined 
by the ratio of the matrix elements
of $J_{5q}$ and $J_q$, one needs to compare the Symanzik expansions
of these two operators. Combining these, I find (as usual for on-shell
matrix elements)\footnote{%
This equation is a hybrid expression in which
$J_{5q}^b$ and $J_5^b$ are lattice operators, rather than operators
in a continuum effective theory, while 
$\bar q \sigma\cdot F\gamma_5 T^bq$ is a
continuum operator, in which the regularization and renormalization
is such that no mixing with lower dimension operators is allowed.
This equation can be obtained by rearranging the Symanzik expansions for the two
operators.}
\begin{equation}
J_{5q}^b = \frac{\mres}{a} J_5^b \left[1 + {\cal O}(a^2m^2)\right]
+ a c_5 \bar q \sigma\cdot F \gamma_5 T^b q 
+ \dots\,.
\end{equation}
I have kept operators up to dimension 5, and dropped all
contributions proportional to $\mres^2$.
This is just eq.~(\ref{eq:J5qtoJ5}) with the $O(a)$ term included.
Note that $\mres$ and $c_5$ are the same quantities
as in eq.~(\ref{eq:Symanzik})---this is
required in order to reproduce the result (\ref{eq:mres'vsmres}) for $\mres'$.

The dominant contribution to $\mres'$ linear in $m_q$ (or, equivalently,
linear in $m$) arises from
the different mass dependences of
{$\langle 0|J_5^b|\pi\rangle$} and
{$\langle 0|\bar q\sigma\cdot F\gamma_5 T^bq|\pi\rangle$}.
The result is a contribution quadratic in $a$,
whose relative size is not suppressed by $\mres$:
\begin{equation}
\mres'(m) = \mres'(m=0) \left[1 + {\cal O}(m a^2 \Lambda) \right]\,.
\end{equation}
Another way of stating this is to consider the derivative of
$\log(\mres')$ with respect to the dimensionless bare quark mass:
\begin{equation}
\frac{1}{\mres'}\frac{d \mres'}{d (am)}\Bigg|_{m\approx -\mres'/a}
\approx
\frac{1}{\mres'}\frac{d \mres'}{d (am)}\Bigg|_{m\approx 0}
= {\cal O}(a \Lambda) \,.
\end{equation}
(Note that the derivative can equally well be evaluated at
$m_q=0$ or $m=0$ at the order I am working.)
Crudely evaluating this derivative from the plots in 
Refs.~\cite{Antonio06,Allton07}
I find that for $1/a=1.6\,$GeV,
the l.h.s. is $\approx 1.7$ for both $L_5=8$ and $L_5=16$.
This implies a scale $\Lambda\approx 3\,$GeV, which,
while very large, is not unheard of. For example, 
$d m_\pi^2/d m_q \approx 5\,$GeV,
and this large scale leads to enhanced discretization errors
in improvement coefficients for improved Wilson fermions~\cite{GuptaImp}.
Nevertheless, the large scale does make one worry that something is
missing from the analysis given here.

Although the slope is larger than expected, it only produces
a small ``uncertainty'' in the value to use for $\mres'(m)$.
A natural choice is to use $\mres'(-\mres'/a)$, i.e. the value at $m_q=0$.
Another commonly used option is $\mres'(0)$. The difference between these
two is of relative size ${\cal O}(\mres' a \Lambda)$, a correction of
less than a percent even if $a\Lambda \approx 2$.
Because of this, in the following I will treat $\mres'$ as a mass-independent
constant.

A striking feature of numerical results for $\mres'$ is that the slope with
respect to the valence quark mass is significantly larger (by a factor
of $\approx -2.5$) than that with respect to the sea quark 
mass~\cite{Antonio06,Allton07}.
The corresponding scale, $\Lambda$, approaches $10\,$GeV. 
Why such a large scale enters is an unresolved puzzle that bears further thought.

\subsection{Enhanced $\mres$ effects: condensate}

The small size of the generic $\chi$SB effects exemplified by
eq.~(\ref{eq:genericerrors}) implies that
problems can only occur if such effects are enhanced.
Two enhancement mechanisms are
power divergences, i.e. mixing with operators of lower dimension,
and ``chiral enhancement'', i.e.
mixing with operators with less suppressed chiral behavior. 
It turns out that the former is much more effective.

As a first example of enhanced $\chi$SB contributions
I consider the quark condensate, $\langle \bar q q \rangle$.
Note that one can also calculate the condensate indirectly,
using $\chpt$ and
results for $f_\pi$, $m_\pi^2$ and quark masses, but it is
also of interest to do a direct, $\chpt$ independent, calculation.

To calculate the condensate one needs, in principle,
to first take the infinite volume limit and then the chiral limit.
Both present significant practical challenges, but I assume here that these
have been overcome. The question I address is whether, after taking
these limits, and in particular after taking $m_q\to 0$, the result
has significant contamination due to the lack of exact chiral symmetry.

This contamination has been considered in Ref.~\cite{RBC00},
and I recall and recast the theoretical analysis given in that work.
Consider the Symanzik expansion of the scalar bilinear,
which is schematically of the form
\begin{equation}
(\bar q q)\Big|_{\rm DWF} \sim 
\frac{m+ x\mres/a}{a^2} + 
(\bar q q)\Big|_{\rm cont.}
+ \dots \,,
\label{eq:Symcond}
\end{equation}
where I am using $\mres$ as defined by the Symanzik expansion,
eq.~(\ref{eq:Symanzik}).
Here I keep only the leading two operators 
(the identity operator and the condensate itself),
and identify schematically only the leading terms in the coefficients
of these operators.
Chiral symmetry requires that the coefficient of the identity operator
is proportional to the quark mass, and thus quadratically divergent,
but $\chi$SB allows a $1/a^3$ divergence proportional to $\mres$.
Note that since UV momenta dominate the mixing, one cannot use
the Symanzik action eq.~(\ref{eq:Symanzik}) inside loops. Thus $m$
and $\mres/a$ do not appear in the linear combination which gives $m_q$.
In other words, instead of $x=1$, one expects $x={\cal O}(1)$.\footnote{%
A heuristic discussion of why this arises goes as follows.
In the quark loop which gives the condensate,
higher order terms in the Symanzik expansion 
(\protect\ref{eq:Symanzik}), i.e. the Pauli term, etc.,
 are not suppressed---overall factors of $a$
are canceled by factors of momenta $k\sim 1/a$.
These higher order terms are not removed by adjusting the bare mass to
cancel the contribution from the leading ($\mres$) term.
One reaches the same conclusion by
comparing expressions for the condensate
and for $\mres$ using the transfer matrix formalism~\cite{Christlat05,RBC07}.}
Thus, if one extrapolates to $m_q=m+\mres/a=0$ one finds
\begin{equation}
\lim_{m_q\to 0} \lim_{L\to\infty} \langle \bar q q \rangle_{\rm DWF}
= \langle \bar q q\rangle_{\rm cont} + (x-1) \frac{\mres}{a^3} + \dots
\end{equation}
The second term is the result of the finiteness of $L_5$. 
Since $x$ is not known, this term gives an uncontrolled error in
the condensate. It can be studied and reduced only by increasing $L_5$---a
very expensive proposition.

How large is the resulting error in the condensate for present simulations?
Since $\langle\bar qq\rangle_{\rm cont} \approx \LQCD^3$ (with a numerical
factor not far from unity), the relative error is
\begin{equation}
\frac{\delta \langle \bar q q\rangle}{\langle \bar qq\rangle}
\sim 
\frac{\mres}{a^3\LQCD^3} 
\,.
\label{eq:conderror}
\end{equation}
This is enhanced by {\em four} powers of $a \LQCD$ compared to
the generic correction exemplified by eq.~(\ref{eq:genericerrors}).
Numerically,
\begin{equation}
\frac{\delta \langle \bar q q\rangle}{\langle \bar qq\rangle}
\approx \frac{3\times 10^{-3}}{(0.3/1.6)^3} \approx 0.5
\label{eq:conderrorcestimate}
\end{equation}
for the present $0.12\,$fm simulations, and only slightly smaller for those at
$0.09\,$fm ($\approx 0.4$ if $\mres\approx 10^{-3}$).
I conclude that one cannot directly calculate the condensate with present
and planned DWF simulations, except to check the order of magnitude.

I stress that these considerations do not apply to the calculations using
overlap fermions, using which several groups have studied the condensate in small
volumes and compared to the $\epsilon-$regime predictions of $\chpt$.
The chiral symmetry breaking term is absent or negligible with overlap fermions.

\subsection{Enhanced $\mres$ effects: $\epsilon'/\epsilon$}

I now turn to a much more important example: the matrix elements
which must be calculated to determine whether the Standard Model predicts
the measured value of $\epsilon'/\epsilon$.
One needs to calculate the $K\to\pi\pi$ matrix elements of the operators
in the $\Delta S=1$ electroweak Hamiltonian, the most important of which is
\begin{equation}
{\cal O}_6 =\bar s_a \gamma_\mu {(1-\gamma_5)} d_b
\sum_{q=u,d,s} \bar q_b \gamma_\mu {(1+\gamma_5)} q_a\,,
\label{eq:O6}
\end{equation}
(where $a$ and $b$ are color indices).
While ultimately a high precision calculation would be desirable,
even a result with 10-30\% precision would be very interesting.

As is well known, many challenges must be overcome to calculate
these matrix elements. One is the difficulty of a direct calculation involving
a two pion final state with physical (or near physical) kinematics---the
method is known in principle, but involves very large lattices and
the need to pick out excited states~\cite{LL}.
This can be circumvented, at the cost of a loss of some precision,
by using $\chpt$ to relate the physical matrix elements to unphysical ones
which are easier to calculate:
\begin{itemize}
\item
At leading order (LO), one can use $K\to 0$ ($m_d\ne m_s$) and $K\to\pi$ ($m_d=m_s$)
matrix elements~\cite{Bernardetal}; 
\item
At next-to-leading order (NLO), one must add further matrix elements to
determine all the needed low-energy constants. One scheme uses
$K\to\pi$ with $m_d\ne m_s$,
$K\to\bar K$ matrix elements of the $\Delta S=2$ Hamiltonian,
and unphysical $K\to\pi\pi$ matrix elements with the pions at 
rest~\cite{LaihoSoni}.\footnote{%
One can also use partially quenched matrix elements to provide further
information and cross-checks. This is discussed in
Refs.~\cite{LaihoSoni,sailing}. For simplicity, I do not discuss
this extension here.}
Another uses $K\to\pi\pi$ matrix elements involving particles
with non-vanishing momenta~\cite{SPQR}.
\end{itemize}
The precision that one can obtain using these two methods is unclear.
Naively, for kaon matrix elements, the relative errors in the LO calculation
are of size $(m_K/\Lambda_\chi)^2 \approx 0.2$, where $\Lambda_\chi=4\pi f$,
while those in the NLO calculation are $\sim (m_K/\Lambda_\chi)^4 \approx 0.04$.
There is considerable evidence, however, that these errors are larger for
$\epsilon'$, and it may require a NLO calculation
to achieve the desired 10-30\% precision.
In practice, the LO calculation will be done first (and is indeed underway now),
with the NLO refinements added subsequently~\cite{Mawhinneytalk}.
The LO methodology has been worked out and tested in quenched 
calculations~\cite{NoakiquenchedME,RBCquenchedME}.

This brings me to the point I want to address here---how is
this calculation affected by the lack of exact chiral symmetry?
This has been discussed in part by Refs.~\cite{NoakiquenchedME,RBCquenchedME},
and more extensively in Ref.~\cite{sailing}, but
here, based on considerations 
with Wilson-like fermions~\cite{Maianietal,Dawsonetal},
I extend the discussion.

Even with exact chiral symmetry, as soon as one considers unphysical matrix
elements one must account for mixing with the lower-dimension
operator~\cite{Bernardetal}\footnote{%
Here and in the following I will assume for definiteness
that the quark masses include the shift due to $\mres'$, i.e.
are given by eq.~(\ref{eq:mqdef}), although one could
equally well use bare lattice quark masses for this analysis.}
\begin{equation}
{\cal O}_{\rm sub1} = \frac{1}{a^2} \left\{ (m_d-m_s) \bar s \gamma_5 d 
                             + (m_s+m_d) \bar s d\right\}
\,.
\end{equation}
In the $\chpt$-based methods discussed above this is subtracted using
the $K\to0$ matrix element. This only works, however, if the positive
and negative parity parts of the operator are related
as in ${\cal O}_{\rm sub1}$, which requires chiral symmetry.
The breaking of chiral symmetry with DWF leads to
additional operators to subtract. I list here those of
dimension less than 6 (which thus have power divergent coefficients
and give enhanced contributions) which are linear in $\mres$
(as opposed to quadratic or higher order)
\begin{eqnarray}
{\cal O}_{\rm sub2} &=& \frac{\mres}{a^3}\bar s d\,,\\
{\cal O}_{\rm sub3} &=& \frac{\mres}{a} \bar s \sigma\cdot F d\,,\\
{\cal O}_{\rm sub4}^{\pm} &=& \frac{\mres (m_s \pm m_d)^2}{a} \bar s d\,,\\
{\cal O}_{\rm sub5} &=& \frac{\mres (m_s^2-m_d^2)}{a} \bar s \gamma_5 d \,.
\end{eqnarray}
Because of the presence of UV divergences,
it is the generic $\mres$ that appears in these operators.
This list is almost the same as for Wilson fermions~\cite{Maianietal},
except that here each occurrence of a ``$L\leftrightarrow R$ flip''
is always suppressed,
either by a factor of $\mres$ or of $m_q$.\footnote{%
In fact, this is not quite correct:
near-zero modes allow multiple flips at the cost of a single $\mres$
suppression~\cite{sailing}.
As explained in the talk by Christ~\cite{Christtalk},
this allows an operator
of the form $(\mres/a^2) m_q \bar s d$, which would naively
require an additional factor or $\mres$. 
I discuss this operator further below.}
Note that the operators ${\cal O}_{\rm sub4}^\pm$ and ${\cal O}_{\rm sub5}$
are usually subsumed into mass-dependent coefficients of 
${\cal O}_{\rm sub1}$ and ${\cal O}_{\rm sub2}$.
I prefer to keep the new operators explicit, since the factors of $m_q$
impact the fits to $\chpt$.

The presence of these operators invalidates
the $\chpt$-based methods, and if not subtracted, they lead to
errors in matrix elements that can only be reduced by increasing $L_5$.
Clearly the greater the power of the UV divergence, the larger the
potential error. 

The presence of the most divergent operator,
${\cal O}_{\rm sub2}$, was noted in the quenched 
studies~\cite{NoakiquenchedME,RBCquenchedME}, and
a method was proposed and implemented to subtract it within the LO $\chpt$
approach. I recall this method briefly here as I will build on it below.
In order to determine $\langle K|{\cal O}_6|\pi\pi\rangle$ at LO in $\chpt$
one needs to calculate, for $m_s=m_d$,
\begin{equation}
{\cal M}_{\rm phys}
= \langle K| {\cal O}_6 - c_1 {\cal O}_{\rm sub1} - c_2 {\cal O}_{\rm sub2} | \pi\rangle
\,,
\end{equation}
where I have dropped ${\cal O}_{\rm sub3}$ etc. for now.
The subtraction coefficient
$c_1$ is to be determined from the $K\to0$ matrix element, and I assume that
this can be done with sufficient accuracy, as seems likely from the quenched
studies. The problem then arises because the
coefficient $c_2$ is not known. Without a determination of $c_2$, the
matrix element will have an unacceptably large error
\begin{equation}
\frac{\delta {\cal M}_{\rm phys}}{{\cal M}_{\rm phys}}
\sim \frac{c_2 \mres B/a^3}{m_K^2\LQCD^2} 
\sim \frac{\mres}{(a\LQCD)^3} \frac{\LQCD}{m_s}
\sim 1
\,.
\label{eq:c2error}
\end{equation}
Here I use the estimate
${\cal M}_{\rm phys} \approx m_K^2 \LQCD^2$,
which incorporates the chiral suppression and
is consistent with numerical results from quenched simulations
and continuum estimates.
I also use the LO $\chpt$ result 
$\langle K|\bar s d|\pi \rangle \approx B \approx m_K^2/m_s$,
and take $c_2\approx 1$, $m_s\approx 90\,$MeV, $\LQCD=300\,$MeV, 
and, finally,
 $\mres=0.003$ for $1/a=1.6\,$GeV, as above. 
Note that the errors are of the same
form as the condensate, 
eq.~(\ref{eq:conderror}),
except for the additional chiral enhancement.

This error due to $c_2\ne 0$
can be removed, however, with a second subtraction condition.
One approach is to enforce that ${\cal M}_{\rm phys}$ vanishes in the chiral limit, 
since this holds in a chirally symmetric theory.
The idea is then to adjust $c_2$ until this vanishing holds. If one works
at LO in $\chpt$ it is then the slope versus $m_s=m_d$ that gives the
required information~\cite{NoakiquenchedME,RBCquenchedME}. 
At NLO a more sophisticated fit to the data is required, including quadratic
terms and $m_s\ne m_d$, but the same subtraction condition works.\footnote{%
One can avoid the need to rely on accurate fitting of the 
matrix element by using the ratio
\begin{equation}
\frac{\langle K| {\cal O}_6 - c_1 {\cal O}_{\rm sub1} | \pi\rangle}
     {\langle K| \bar s d| \pi\rangle}
=
\frac{{\cal M}_{\rm phys}}{\langle K|\bar s d| \pi \rangle}
+ c_2 \frac{\mres}{a^3}
\,,
\end{equation}
The first term on the r.h.s. has a chiral expansion beginning at linear order
in $m_q$, so it should be straightforward to determine and remove the
$c_2$ term, which is a constant.
One can then reconstruct ${\cal M}_{\rm phys}$ by multiplying by
the separately measured scalar matrix element ${\langle K|\bar s d| \pi \rangle}$.
This method is due to the RBC collaboration.}

If ${\cal O}_{\rm sub2}$ can be subtracted,
then the largest uncertainty is likely due 
to the next-most divergent operator ${\cal O}_{\rm sub3}$.
The presence of this operator is well known to Wilson and DWF practitioners,
but it is only now, with the advent of realistic DWF simulations,
that it becomes relevant to estimate its contribution.
Since its $K\to\pi$ matrix element does not vanish in the chiral limit,
it contaminates the subtraction method for ${\cal O}_{\rm sub2}$---a
single condition (vanishing in the chiral limit) cannot determine
both unknowns $c_2$ and $c_3$.
One must either accept the resulting error or determine another subtraction
method. The former approach may, in fact, be sufficient.
One expects that the ratio of matrix elements of $\bar s \sigma\cdot F d$ 
and $\bar s d$ is approximately $\LQCD^2$.
Using this, and the parameters given above,  one finds 
\begin{equation}
\frac{\delta {\cal M}_{\rm phys}}{{\cal M}_{\rm phys}}
\sim \frac{c_3 \mres}{a\LQCD} \frac{\LQCD}{m_s} 
\sim \frac{\mres}{a m_s}
\sim 0.05
\label{eq:c3error}
\end{equation}
for the $a=0.12\,$fm simulations.
At $a=0.09\,$fm, the result is 2-3 times smaller.
Thus the error one makes by ignoring this term is comparable to, or smaller
than, that due to the missing
next-to-next-to-leading (NNLO) contributions 
in the $\chpt$ analysis. It seems to me that one can, at this stage,
proceed by ignoring this operator.

It should also be possible to subtract the contributions of
this operator. One idea (which I have not thought through in detail)
is to use the fact that
${\cal M}_{\rm phys}$ is proportional to $m_\pi m_K$, whereas the 
matrix element of ${\cal O}_{\rm sub3}$
also has terms of the form $m_\pi^2$ and $m_K^2$
(which are non-leading for this operator but will contribute
to the putative ${\cal M}_{\rm phys}$ if ${\cal O}_{\rm sub3}$
is not subtracted).
The idea would then be to
adjust $c_3/c_2$ so as to remove these terms.
Another approach is suggested in Ref.~\cite{Maianietal}.
Both ideas are valid only at LO in $\chpt$ and would need to
be elaborated in order to work at NLO.

I now return to the remaining operators in the list,
${\cal O}_{{\rm sub4}^\pm}$ and ${\cal O}_{\rm sub5}$.
Given the explicit factors of quark mass, these do not affect the LO
$\chpt$ analysis, but do contaminate the NLO analysis.
They lead to an error which is suppressed by $(a m_s)^2$ relative to that
which would be present if ${\cal O}_{\rm sub2}$ were not subtracted
[see eq.~(\ref{eq:c2error})], and thus of size
\begin{equation}
\frac{\delta {\cal M}_{\rm phys}}{{\cal M}_{\rm phys}}
\sim \frac{c_{4,5} \mres}{(a\LQCD)^3} \frac{\LQCD}{m_s} (a m_s)^2
\sim \frac{\mres}{a m_s} \frac{m_s^2}{\LQCD^2}
\sim 0.005
\,.
\end{equation}
It appears, therefore, that one can safely ignore these operators
for a long time to come.

The discussion thus far turns out to be incomplete---there is an additional
contribution noted by Norman Christ in his talk here~\cite{Christtalk}.
This arises from an operator which is naively be suppressed by $\mres^2 m_q$,
but is in fact
suppressed only by $\mres m_q$ because a single near-zero mode 
can simultaneously flip the chirality
of two quarks as long as they have different flavors
(and the chirality flips are in the ``same direction'').
The fact that the zero-modes are localized causes no suppression
because the initial operator itself is local.\footnote{%
This is in distinction to the generic contributions from
the near-zero modes, which, as explained in Ref.~\cite{sailing},
are suppressed by additional factors of $a$ due to the localized
nature of the eigenmodes.}
Christ shows that one obtains the operator
\begin{equation}
\frac{\mres}{a^2} \left\{
(m_s-m_d) \bar s \gamma_5 d + (m_s + m_d + 2 m_u) \bar s d \right\} \,.
\end{equation}
Part of this operator is removed by the subtraction for ${\cal O}_{\rm sub1}$, 
allowing one to rewrite it as
\begin{equation}
{\cal O}_{\rm sub6} = \frac{\mres}{a^2} 2 (m_s+m_d+m_u) \bar s d
\,.
\label{eq:ONC}
\end{equation}

How does the error due to this operator compare with that
from ${\cal O}_{\rm sub3}$? There are competing factors here.
On the one hand ${\cal O}_{\rm sub6}$ is enhanced by one power of
$1/a$, but, on the other hand, it is chirally suppressed by one factor of $m_s$.
Naively combining these factors one obtains an enhancement of
$m_s/(a\LQCD^2)\approx 2$. There is another suppression factor, however,
namely that it is only the near-zero mode part of $\mres$, 
i.e. the second term in eq.~(\ref{eq:mresform}), which contributes.
As noted above (footnote 5) the near-zero mode contribution to
$\mres'$ is, for present simulations, at most comparable to that from the
extended modes. Thus it
seems plausible that the $\mres$ appearing in (\ref{eq:ONC}) is smaller
by at least by a factor of 2 than that appearing in
the error estimate due to ${\cal O}_{\rm sub3}$.
All in all, I expect that ${\cal O}_{\rm sub3}$ and ${\cal O}_{\rm sub6}$
give comparable errors, and thus, based on the
estimates given above, that both can probably be ignored at
this stage of the numerical calculations.

Eventually, however, one will want to subtract ${\cal O}_{\rm sub6}$ as well.
As noted in Ref.~\cite{Christtalk}, this is possible,
in principle, by varying $m_u$ independently from $m_d$ and $m_s$.

In conclusion, as long as one implements a method to subtract the most
divergent operator mixing caused by $\chi$SB, DWF can probably be used
to study $\epsilon'/\epsilon$ using $\chpt$-based methods
(particularly the NLO method)
with sufficient accuracy to make significant progress.

\subsection{Implications of $\mres$ for $B_K$}

I now turn to another phenomenologically important matrix element,
$B_K$. This is an example where 
there is no mixing with lower dimension operators,
but mixing with chirally enhanced operators can occur.
Thus one might expect a ``chiral enhancement'' of \CSB\ effects.
It is also a quantity where percent level precision is highly desirable,
since the better we can calculate $B_K$, the more precisely we can
test the Standard Model. 

Chiral enhancement can occur through mixing of the original $B_K$
operator, ${\cal O}_K=\bar s_L\gamma_\mu d_L \bar s_L \gamma_\mu d_L$,
with ``L-R'' operators exemplified by
$\bar s_L\gamma_\mu d_L \bar s_R\gamma_\mu d_R$. The L-L operators
have $K-\bar K$ matrix elements which are chirally suppressed, and
thus proportional to $m_s$, while the L-R operators are unsuppressed.
Since two quarks must have their chirality flipped, mixing at
the perturbative level is proportional to $\mres^2$ (the argument
being identical to that for $Z_A-1$ discussed above).\footnote{%
Note that near-zero
modes do not give a contribution proportional to $\mres$ here, just
as for $Z_A-1$~\cite{Christlat05}.}
Even with a chiral enhancement of size $\LQCD/m_s$,
this contribution is numerically insignificant. 
Including discretization errors [i.e. working out the
Symanzik expansion for ${\cal O}_K$ and keeping terms
analogous to the Pauli-term in 
(\ref{eq:Symanzik})]
I find that there is a mixing proportional to $\mres a m_s$.
The factor of $m_s$ is needed because of chirality---one flip requires
$a \mres$ and the other $m_s$. Here $\mres$ is meant generically,
since the contribution involves logarithmically divergent integrals
and thus UV momenta.
The extra factor of $m_s$ in the mixing cancels the chiral enhancement
in the matrix element, so one finds in the end an error which
has the same size as those in generic hadronic quantities, i.e.
\begin{equation}
\frac{\delta B_K}{B_K} \sim \mres a m_s \frac{\LQCD}{m_s}
\sim \mres a \LQCD \approx (2-9) \times 10^{-4} \,.
\end{equation}
Here the numerical estimate is from eq.~(\ref{eq:genericestimate}).
Errors of this same size also arise directly from the Pauli-term
in the Symanzik Lagrangian.

I conclude that \CSB\ is not a concern for percent level calculations
of $B_K$ with DWF. Indeed, such precision is already being attained
at a single lattice spacing~\cite{DWFBK}, with a second lattice spacing
underway~\cite{Mawhinneytalk}.

\subsection{Pion EM splitting}

I close with a final example which is both of phenomenological
interest, and has particular resonance in this meeting as
it was based on a suggestion (long ago) by David Kaplan.
The idea is to calculate the pion EM mass splitting using
$\chpt$ to relate it to a simpler quantity:
\begin{equation}
\Delta_\pi \equiv \frac{m_{\pi^+}^2 - m_{\pi^0}^2}{2 e^2/f_\pi^2}
= - 
\int d^4x \frac{1}{4\pi^2x^2} 
\langle \bar u_L \gamma_\mu d_L(x) \ \bar d_R \gamma_\mu u_R(0)\rangle
\,,
\label{eq:Das}
\end{equation}
where the r.h.s. should be extrapolated to the chiral limit.
The r.h.s. is part of the vacuum energy due to electromagnetism,
with the $1/x^2$ arising from the photon propagator.
This application of (what was then called) current algebra and PCAC
was suggested in 1967 by Ref.~\cite{Dasetal}, and they evaluated
the correlation function using resonance saturation constrained by
the Weinberg sum rules. For an updated discussion incorporating
large $N_c$ input, see, e.g., Ref.~\cite{Peristalk}

Of course one can also do a direct calculation of EM splittings,
using a background EM field, and indeed such calculations
are underway using DWF~\cite{EMsplit}. But it would be interesting to have
a result using a different method, and also to study correlation
functions like this so as to make contact with large-$N_c$ based
approaches discussed here by Golterman~\cite{Maartentalk}.

A first lattice calculation (which happens also to be
my first lattice calculation) was
attempted long ago, using Wilson fermions on a lattice of size too small to
mention~\cite{GKS84}. We understood at the time
that chiral symmetry was essential
for evaluating the r.h.s. of (\ref{eq:Das})---it is a chiral order parameter,
like the condensate, since its L-R structure means that
it vanishes in perturbation theory to all orders for massless quarks.
It quickly became clear that the $\chi$SB inherent to Wilson fermions 
introduced an intolerable error.
Given the much improved chiral symmetry of DWF, I thought it interesting
to revisit this quantity and see how well one can do.

To study this, I note that, if $\mres=0$, one expects, schematically,
that
\begin{equation}
\Delta_\pi \approx f_\pi^4 + \frac{m_q^2}{a^2} + m_q \langle \bar qq\rangle
\ln a+ \dots
\,,
\end{equation}
where the first term is the desired result coming from long-distance 
contributions to the integral over $x$
(with $f_\pi^4$ being a good estimate of its magnitude
using the experimental pion splitting), 
while subsequent terms are UV
divergent and can be determined using the operator product expansion.
(In doing so, I have not kept track of factors of $\alpha_S$ in
coefficient factors, since I am making rough estimates.)
As in the case of the condensate, one can determine the order of magnitude of
the contribution due to $\mres$ by substituting $m_q \to \mres/a$.
In this way one finds that, after chiral extrapolation, the relative error
is
\begin{equation}
\frac{\delta \Delta_\pi}{\Delta_\pi}
\sim
\frac{\mres^2}{(a f_\pi)^4} + \frac{\mres}{a f_\pi}\times 
\frac{\langle \bar qq \rangle}{f_\pi^3}
\sim
{\cal O}(1) + {\cal O}(1)
\,.
\end{equation}
In words, both contributions quadratic and linear in $\mres$ turn out,
for present and near-future DWF simulations, to give errors of order 100\%.
I conclude that it is probably not practical to calculate $\Delta_\pi$
in this way,
but it is perhaps worth a more detailed look.

\section{Conclusions}

It appears that the residual breaking of chiral symmetry is problematic
only for highly UV divergent quantities: in the examples I have 
considered, these are the
condensate and the pion EM mass splitting.
Thus I take the middle path and conclude that DWF are
{\bf W}ONDERFUL:
there is no practical barrier to calculating
{\bf many} quantities of interest (spectrum, $B_K$ and related matrix elements,
nucleon properties, \dots, and
possibly $\epsilon'/\epsilon$)
with desired precision in the next five years.
This will require, however, simulations with parameters extending to
at least $m_\ell/m_s\approx 0.1$ and 
$L \approx 4\,$fm, 
and probably down to $a\approx 0.06\,$fm,
while keeping  $\mres\approx 10^{-3}$.
Given present estimates of CPU requirements, these appear to be
attainable parameters in a five year time-scale.

Given the exciting potential of DWF simulations, I cannot resist
an exhortation.
For an outsider, a striking {\em lacuna} in present DWF studies is
the lack of calculations of quantities involving heavy (and in particular
$b$) quarks.
There are so many interesting and important heavy-light
quantities to calculate that this is a serious omission.
More generally, as stressed here by Soni, the lattice community
now has a great new method, and
needs to think hard about increasing its repertoire.

Clearly, the next such DWF meeting should be very interesting!

\section*{Acknowledgments}
\label{sec:acknowledge}

I thank Tom Blum and Amarjit Soni for the invitation to speak and for
organizing a very stimulating workshop.
I thank Norman Christ, Meifeng Lin and
particularly Maarten Golterman and Yigal Shamir
for very helpful correspondence and discussions.
This work was supported in part by the US Department of Energy.

\end{document}